\documentclass[prd,twocolumn,nofootinbib,showpacs,aps]{revtex4-1}

\usepackage{natbib}

\usepackage{amsfonts}
\usepackage{amsmath}
\usepackage{amssymb}
\usepackage{bm}
\usepackage{dcolumn}
\usepackage{graphicx}
\usepackage{graphics}
\usepackage[latin1]{inputenc}
\usepackage{latexsym}
\usepackage{rotating}
\usepackage{hyperref}
\usepackage[all]{hypcap} 
\usepackage{xspace} 
\usepackage[usenames]{color}
\usepackage{mathrsfs}

\usepackage{ulem}
\definecolor {darkgreen}{rgb}{0.2,0.7,0.2}

\newcommand\be{\begin{equation}}
\newcommand\bea{\begin{eqnarray}}
\newcommand\ee{\end{equation}}
\newcommand\eea{\end{eqnarray}}
\newcommand\bw{\begin{widetext}}
\newcommand\ew{\end{widetext}}

\newcommand{\nn}{\nonumber}

\newcommand{\p}{\partial}
\newcommand{\D}{\mathcal{D}}

\begin{document}
\title{Multipole moments in scalar-tensor theory of gravity}    

\author{George Pappas$^{1}$ and Thomas P. Sotiriou$^{1,2}$}
\affiliation{$^{1}$School of Mathematical Sciences,  University of Nottingham,
University Park, Nottingham NG7 2RD, UK\\
$^{2}$School of Physics and Astronomy, University of Nottingham,
University Park, Nottingham NG7 2RD, UK}

\begin{abstract} 
Stationary, asymptotically flat spacetimes in general relativity can be characterized by their multipole moments. The moments have proved to be very useful tools for extracting information about the spacetime from various observables and, more recently, for establishing universalities in the structure of neutron stars. As a first step toward extending these methods beyond general relativity, we develop the formalism that allows one to define and calculate the multipole moments in scalar-tensor theories of gravity.  
\end{abstract}

\date{\today}
\maketitle

\section{Introduction}

General relativity (GR) has been tested quite successfully in various regimes and astrophysical conditions \cite{will-living} and is, therefore, the established theory of gravity. Despite GR's success, various alternative theories of gravity have been proposed (see for example Ref.~\cite{Clifton2011jh}).
The motivation for considering alternative theories is threefold: i) one might hope for an improved behavior in the ultraviolet, as GR is a nonrenormalizable theory; ii) alternative theories might accommodate mechanisms that can account for either dark energy or dark matter; iii) consistent alternative theories are needed in order to systematically check whether observations indeed show a preference for GR.

While terrestrial, solar system and cosmological experiments and observations place stringent constraints on the gravitational interaction, they only test the weak field regime. The next frontier is to test gravity in the strong field regime, where deviations from the standard picture could be larger. 
Black holes are considered to be the most suitable candidates for testing deviations from GR in this regime, mainly because of their simple structure. Compact stars, and in particular neutron stars (NSs) are much harder to model because there is still significant uncertainty about their internal structure. The behavior of matter at densities that exceed that of a nucleus is still somewhat unknown and this is reflected  in an ambiguity in choosing the right equation of state (EOS). The uncertainty in the EOS leads to uncertainties in the spacetime structure and the properties of the star, and these uncertainties are comparable to the deviations that considering an alternative theory of gravity would introduce.

This picture has started to change though,  due to recent evidence that certain relations between specific quantities that can be used to characterize NSs  are universal, i.e., these relations are insensitive to the particular realistic EOS one might use to calculate the structure of the NS. These universal relations fall under two major categories, those that are related to dynamical properties of NSs in binary inspirals and are observed in the spectrum of gravitational waves (see for example Refs.~\cite{Read.et.al.2013PRD,Bernuzzi2014PRL,Rezzolla2014PRL,Chan:arXiv1408.3789}), and those that are related to the properties of the structure of isolated NSs. The latter category can be further separated into two subcategories. The first one comprises the so-called {\it I-Love-Q} relations introduced by Yagi and Yunes, that relate the moment of inertia ($I$), the rotationally induced quadrupole ($Q$), and the quadrupolar tidal deformability ({\it Love} number, $\lambda$) \cite{YY2013Sci,YY2013PhRvD} (there have been many follow-up investigations of these relations for rapidly rotating and magnetized NSs, as well as inspiraling NSs \cite{Maselli2013PRD,Haskel2014MNRASL,Doneva2014ApJL,Pappas:2013naa,Chakrabarti2014PRL,Martinon2014PRD}). The second comprises the so-called three-hair relations that relate the higher multipole moments of the spacetime of a NS to the first three nonzero multipole moments, i.e., the mass ($M$), the angular momentum ($J\equiv S_1$), and the quadrupole moment ($Q\equiv M_2$). This type of relations was first introduced for realistic EOSs in full GR by Pappas and Apostolatos for the spin octupole ($S_3$) \cite{Pappas:2013naa} and was later extended to the mass hexadecapole ($M_4$) \cite{Yagi:2014bxa}. This type has also been investigated in Newtonian theory \cite{Stein2014ApJ,Katerina2014PhRvD} (recently an investigation on why this type of relations is satisfied has also been presented \cite{yiloveq}).    

Upon their discovery, it was proposed that such relations could be used to test different theories of gravity since they may differ from the one theory to the other, providing thus a distinguishing characteristic. In particular the {\it I-Love-Q} relations have already been investigated in various proposed modifications of GR, such as dynamical Chern-Simons \cite{YY2013Sci}, Eddington-inspired Born-Infeld gravity \cite{EiBI2014ApJ}, Einstein-Gauss-Bonnet-dilaton theory of gravity \cite{dEGB2014PhRvD}, and scalar-tensor theory both for slowly rotating \cite{PaniBerti2014PhRvD} and rapidly rotating NSs \cite{Doneva2014PhRvD}. So far, the only theory that has been shown to have {\it I-Love-Q}  relations that are distinguishable from those in GR is dynamical Chern-Simons theory of gravity, while all the others appear to have the same behavior as GR. This can be considered disappointing or intriguing, depending on the perspective. Disappointing because {\it I-Love-Q} relations do not seem to be as useful as one would have hoped in distinguishing different theories of gravity; intriguing since it appears that {\it I-Love-Q} relations are universal in a wider class of theories of gravity and that could prove to be very useful for measuring different properties of NSs independently of the particular theory of gravity. 
  
While there have been several investigations of the {\it I-Love-Q} relations in alternative theories, the relations between the multipole moments have not been investigated so far. The reason for this is that while in GR the multipole moments of an asymptotically flat spacetime are well defined, such a definition for alternative theories of gravity does not exist. One of the most thoroughly studied alternatives to GR is scalar-tensor theory \cite{FujiiMaeda,CapozzielloFaraoni,Damour,Jordan49,fierz56,Jordan59,BransDicke,Dicke62}, which can be described by the following action
\begin{equation}
\label{staction}
S=\int d^4x \sqrt{-\hat{g}} \left(\varphi \hat{R}-\frac{\omega(\varphi)}{\varphi} \hat{\nabla}^\mu \varphi \hat{\nabla}_\mu \varphi\right)+S_m(\hat{g}_{\mu\nu},\psi)\,,
\end{equation}
where $\hat{g}$ is the determinant and $\hat{R}$ is the 
Ricci scalar of the metric $\hat{g}_{\mu\nu}$,  
$\hat{\nabla}_\mu$ denotes the corresponding covariant 
derivative, $S_m$ is the matter Lagrangian, and $\psi$ 
collectively denotes the matter fields.  It is  assumed 
that matter fields couple minimally to
$\hat{g}_{\mu\nu}$. 
The aim of the present work is to provide a definition for the multipole moments in the scalar-tensor theory of gravity that will consequently be used to investigate the properties of NS multipole moments in this theory, in the hope of providing a tool for distinguishing scalar-tensor gravity from GR. It is worth pointing out that scalar-tensor theory exhibits very interesting phenomenology related to NSs. Of particular interest is the phenomenon of spontaneous scalarization \cite{Damour93PRL,Damour96PRD} observed in NSs, that can lead to drastically different NS structure in theories that are very close to  GR in the weak field regime.

In the following sections we will first provide an introduction to the definition of multipole moments in GR in the case of a stationary spacetime that admits a timelike Killing field, and then we will extend this definition in the case of the scalar-tensor theory with one massless scalar field. We will then discuss the properties of stationary and axisymmetric solutions in scalar-tensor theory and their formulation in terms of an Ernst potential, and we will provide the formalism for calculating the moments in terms of that Ernst potential. Finally, we will discuss the astrophysical relevance of our result and present our conclusions.

\section{Multipole moments formalism for stationary spacetimes}
\label{sec:moments1}

In the early 1970s, in a series of papers \cite{Geroch70I,Geroch70II,hansen}, a formalism for calculating the multipole moments of asymptotically flat spacetimes that admit a timelike Killing field in general relativity was introduced by Geroch and Hansen. In this context, the multipole moments were defined as tensors at infinity that are generated from a set of appropriate potentials. In the general case, there is one potential for mass moments, and there is one potential for rotation moments. These potentials satisfy field equations on the 3-space of trajectories of the timelike Killing field and encode all the information for the gravitational field.   

A manifold $(S,h_{\mu\nu})$ is considered to be asymptotically flat if there exists a manifold $(\tilde{S},\tilde{h}_{\mu\nu})$ such that,

\begin{enumerate}
\item $\tilde{S}=S\cup\Lambda$, where $\Lambda$ is a single point at infinity
\item $\tilde{h}_{\mu\nu}=\Omega^2h_{\mu\nu}$ is a smooth metric on $\tilde{S}$, and
\item $\Omega|_{\Lambda}=0$, $\tilde{\mathcal{D}}_a\Omega|_{\Lambda}=0$, and $\tilde{\mathcal{D}}_a\tilde{\mathcal{D}}_b\Omega|_{\Lambda}=2\tilde{h}_{ab}$, where $\tilde{\mathcal{D}}_a$ is the covariant derivative associated with the metric $\tilde{h}_{ab}$. 
\end{enumerate}
An example of an asymptotically flat space in the above sense is the Euclidean 3-space, if one chooses the conformal factor $\Omega=r^{-2}$. 

Once the three-dimensional manifold of trajectories of the Killing field is defined and it is conformally extended to include infinity (as long as the manifold is asymptotically flat in the above sense), one can define the multipole moments of the mass and rotation potentials, $\Phi_M$ and $\Phi_J$ respectively, using the following algorithm. First one defines the potential at infinity $\tilde{\Phi}=\Omega^{-1/2}\Phi$, and then the moments are calculated recursively as a set of tensor fields $P_{a_1\ldots a_s}$ from the relations, 
\bea  P &=& \tilde{\Phi},\nn\\
         P_a &=& \tilde{D}_a P,\nn\\
&\vdots& \\
         P_{a_1 \ldots a_{s+1}} &=& \mathbf{\mathcal{C}}\left[\tilde{D}_{a_1}P_{a_2 \ldots a_{s+1}}-\frac{s(2s-1)}{2}\tilde{R}_{a_1a_2}P_{a_3 \ldots a_{s+1}}\right],\nn
  \eea       
where $\mathbf{\mathcal{C}}$ is the symmetric and trace-free operation, and the Ricci tensor, $\tilde{R}_{ab}$, and covariant derivative, $\tilde{D}_{a}$, are those associated with the conformal metric at infinity of the 3-space, $\tilde{h}_{ab}$. The $2^s$ multipole moment is then defined to be the value of $P_{a_1\ldots a_s}$ at $\Lambda$.
   
In what follows we will give a brief description of the setup for the definition of the moments in general relativity for a stationary spacetime and then we will extend this formalism to scalar-tensor theories of gravity with one massless scalar field.  

\subsection{Multipole moments in general relativity}

Consider a source-free solution of Einstein's field equations, i.e., $(M,g_{\mu\nu})$ for which $R_{\mu\nu}=0$. Suppose there is also a timelike Killing field $\xi^{\mu}$. Einstein's field equations for such a spacetime can be written as a set of field equations that involve two scalar fields on a three-dimensional curved space $S$ and the  Ricci tensor $\mathcal{R}_{ab}$ that describes $S$.  One can arrive at these equations as follows.
The norm and the twist of the field are given as 
\begin{eqnarray}
\label{definition1}
\lambda &=& \xi^a \xi_a,\\
\label{definition2}
\omega_a &=& \epsilon_{abcd} \xi^b \nabla^c \xi^d.
\end{eqnarray}
The projection operator and the covariant derivative on $S$ are given by the equations 
\begin{eqnarray}
h_{ab}=g_{ab}-\lambda^{-1}\xi_a\xi_b,\\
h^{ab}=g^{ab}-\lambda^{-1}\xi^a\xi^b,\\
h^a_b=\delta^a_b-\lambda^{-1}\xi^a\xi_b,\\
\mathcal{D}_aT^{bc}=h^k_a h^b_m h^c_n \nabla_k T^{mn},\\
\mathcal{D}^2=\mathcal{D}^a\mathcal{D}_a.
\end{eqnarray}

Starting from the definitions of $\lambda$ and $\omega_a,$ and by using the first Gauss-Codazzi equation, the properties of the Killing field and the projection of the Riemann tensor on $S$, the equations for $\lambda, \omega_a,$ and $\mathcal{R}_{ab}$ that one gets are \cite{GerochTransI},
\begin{eqnarray}
\mathcal{D}^2\lambda&=&\frac{1}{2}\lambda^{-1} (\mathcal{D}^m\lambda) (\mathcal{D}_m \lambda)-\lambda^{-1}           \omega^m\omega_m \nn\\
&&- 2R_{mn}\xi^m\xi^n, \\ 
\mathcal{D}_{[a}\omega_{b]}&=&-\epsilon_{abmn}\xi^m R^n_p \xi^p, \label{twistcond}\\
\mathcal{D}^a\omega_{a}&=&\frac{3}{2}\lambda^{-1}\omega_m\mathcal{D}^m\lambda,\\
\mathcal{R}_{ab}&=&\frac{1}{2}\lambda^{-2}[\omega_a\omega_b-h_{ab}\omega^m\omega_m]+\frac{1}{2}\lambda^{-1}\mathcal{D}_a\mathcal{D}_b\lambda\nn\\
&&-\frac{1}{4}\lambda^{-2} (\mathcal{D}_a\lambda) (\mathcal{D}_b \lambda)+h^m_a h^n_b R_{mn}.
\end{eqnarray}

In the case that there are no sources one has $R_{mn}=0$, the second equation implies that $\omega_a$ is curl-free and, thus, it can be expressed as the gradient of a potential $\omega_a=\nabla_a\omega$. 

In Ref.~\cite{hansen}, instead of the 3-space, $S$, used by Geroch \cite{Geroch70II,GerochTransI}, a conformally related 3-space, $\tilde{S}$, was used in order to construct the field equations for the fields $\Phi$ that will generate the moments. This 3-space is described by the metric, 
\begin{eqnarray}
\tilde{h}_{ab}&=&(-\lambda)h_{ab}=-\lambda g_{ab}+\xi_a\xi_b\,,\\
\tilde{h}^{ab}&=&(-\lambda)^{-1}h^{ab}=-\lambda^{-1} g^{ab}+\lambda^{-2}\xi^a\xi^b\,,\\
\tilde{h}^a_b&=&h^a_b=\delta^a_b-\lambda^{-1}\xi^a\xi_b\,.
\end{eqnarray}
In the way it has been defined, $\lambda$ is negative. We find it more convenient to replace $\lambda$ with $-\lambda$ in all expressions and take it to be positive. 
The conformally transformed metric will then have the form $\tilde{h}_{ab}=\lambda g_{ab}+\xi_a\xi_b$.
The Ricci tensor and the field equations for $\lambda$, and $\omega$ then take the form
\bea
\tilde{\D}^2\lambda&=&\lambda^{-1}\left((\tilde{\D}^m\lambda)(\tilde{\D}_m\lambda)-(\tilde{\D}^m\omega)(\tilde{\D}_m\omega)\right),\\
\tilde{\D}^2\omega&=&2\lambda^{-1}(\tilde{\D}^m\lambda)(\tilde{\D}_m\omega),\\
\tilde{\mathcal{R}}_{ab}&=&\frac{1}{2}\lambda^{-2}\left[(\tilde{\mathcal{D}}_a\lambda)(\tilde{\mathcal{D}}_b\lambda)+(\tilde{\mathcal{D}}_a\omega)(\tilde{\mathcal{D}}_b\omega) \right].
\eea
These equations and the Ricci tensor of the 3-space encode all the information of the Einstein field equations for the spacetime.

Using the scalar quantities $\lambda$ and $\omega$, one can construct the potentials,
\bea \Phi_M &=& \frac{1}{4}\lambda^{-1}(\lambda^2+\omega^2-1),\\
         \Phi_J &=& \frac{1}{2}\lambda^{-1}\omega.
 \eea         
From the field equations for $\lambda$ and $\omega$, one can show that these potentials satisfy a set of ``generalized Laplace'' equations on a curved 3-space, which are equivalent to the initial set of equations for $\lambda$ and $\omega$, and which are of the form

\be\label{laplace}
\mathcal{D}^a\mathcal{D}_a \Phi- (\mathcal{R}/8)\Phi=(15/8)\kappa^4 \Phi,
\ee
where $\Phi$ denotes both $\Phi_M$ and $\Phi_J$, $\mathcal{R}$ is the Ricci scalar of the 3-space, and $\kappa$ is a function that depends on the scalar potentials, 
\be \kappa^4=\frac{1}{2}\lambda^{-2}\left[(\mathcal{D}^m\lambda)(\mathcal{D}_m\lambda)+(\mathcal{D}^m\omega)(\mathcal{D}_m\omega) \right], \ee
and that can be shown to be a smooth function on $\tilde{S}$ and at $\Lambda$ (we have dropped the tilde in the notation). In fact, the notion of the multipole moments of the $\Phi$ potentials depends on the behavior of $\Phi$ at infinity and in particular on it being a smooth function. Therefore, it depends on the conformal structure of the field equations. Since $\Phi$ satisfies an elliptic differential equation, it will be smooth as long as all the coefficients of the equation are smooth.
It has been argued in Ref.~\cite{hansen} that $\kappa$ and the combination $\lambda^{-1}(\lambda^2+\omega^2+1)$ are smooth and that this is enough to imply that $\Phi$ is smooth. 
Hence, one can define the multipole moments from the potentials $\Phi_M$ and $\Phi_J$ using the recursive algorithm presented previously.

\subsection{Multipole moments in scalar-tensor theory of gravity}
\label{sec:moments3}

In action (\ref{staction}) the scalar field $\varphi$ is nonminimally coupled to gravity and has a noncanonical kinetic term.  This representation of the theory is called the Jordan frame. It is far more convenient for our purpose to use the, so-called, Einstein frame representation. The conformal transformation $g_{\mu\nu}=16 \pi G\, \varphi 
\, \hat{g}_{\mu\nu}$, together with the scalar field redefinition
\be
d\phi=\sqrt{\frac{2\omega(\varphi)+3}{4}} \, d\ln\varphi\,,
\ee
will bring action (\ref{staction}) into the following form:
\begin{equation}
\label{stactionein}
S=\frac{1}{16\pi G}\int d^4x \sqrt{-g} \left(R-2 \nabla^\mu \phi \nabla_\mu \phi\right)+S_m(\hat{g}_{\mu\nu},\psi)\,,
\end{equation}
where the matter fields still couple minimally to $\hat{g}_{\mu\nu}$. This implies that $\phi$ is now coupled to the matter fields, and it is this coupling that encodes any deviation from standard GR with a minimally coupled scalar field. 

The advantage of using the Einstein frame in the definition of the moments should now be evident: since for the definition of the moments one needs to work outside the sources, matter can be neglected. Hence, in the Einstein frame one just has to deal with GR with a minimally coupled scalar field. The corresponding field equations outside the sources are
\begin{eqnarray}
\label{feq1}
R_{ab} &=& 2 \partial_a\phi\partial_b\phi,\\
g^{ab}\nabla_a\nabla_b\phi &=&0. \label{feq2}
\end{eqnarray}  
The main difference from standard GR is that now the scalar field acts as a source, that the Ricci tensor does not vanish, and that one is left with one extra equation for the scalar. 

Let us now assume that the metric $g_{\mu\nu}$ admits a timelike Killing vector field $\xi^a$, for which we have $\pounds_{\xi}g_{ab}=0$ and that the scalar field respects this symmetry as well, {\em i.e.}~$\pounds_{\xi}\phi=\xi^a\nabla_a\phi=0$.\footnote{Note that, as long as the scalar field $\varphi$ (or $\phi$) respects the symmetries of $\hat{g}_{\mu\nu}$ , then $\hat{g}_{\mu\nu}$ and $g_{\mu\nu}$ will have the same Killing vectors.} One can then define the norm and the twist of the Killing vector using Eqs.~(\ref{definition1}) and (\ref{definition2}) as before and express the field equations as equations on a three-dimensional hyperspace $S$. The fact that $\xi^a\nabla_a\phi=0$ implies that the twist of the Killing vector is curl-free and thus the potential $\omega$ can be defined as before. Hence, the equations for the potentials $\lambda$ and $\omega$ in terms of Hansen's formalism can take the form 
\bea
\tilde{\D}^2\lambda&=&\lambda^{-1}\left((\tilde{\D}^m\lambda)(\tilde{\D}_m\lambda)-(\tilde{\D}^m\omega)(\tilde{\D}_m\omega)\right),\\
\tilde{\D}^2\omega&=&2\lambda^{-1}(\tilde{\D}^m\lambda)(\tilde{\D}_m\omega),\\
\tilde{\D}^2\phi&=&0,
\eea
where we have also the equation for the scalar field, and finally the equation for the Ricci tensor is
\be
\tilde{R}_{ab}=\frac{1}{2\lambda^{2}}\left[(\tilde{\mathcal{D}}_a\lambda)(\tilde{\mathcal{D}}_b\lambda)+(\tilde{\mathcal{D}}_a\omega)(\tilde{\mathcal{D}}_b\omega) \right]+2(\tilde{\D}_a\phi)(\tilde{\D}_b\phi).
\ee

Using these equations, again one can show that the potentials,
\bea \Phi_M &=& \frac{1}{4}\lambda^{-1}(\lambda^2+\omega^2-1),\\
         \Phi_J &=& \frac{1}{2}\lambda^{-1}\omega,\\
         \Phi_{\phi} &=& \phi,
 \eea 
satisfy, similar to the GR case, generalized Laplace equations with nice conformal properties, the difference being that in this case we have the mass and angular momentum potentials satisfying an equation of the form  
\be\label{laplace2}
\mathcal{D}^a\mathcal{D}_a \Phi- (\mathcal{R}/8)\Phi=\kappa_1^4 \Phi,
\ee
where
\bea \kappa_1^4&=&\frac{15}{16}\lambda^{-2}\left[(\mathcal{D}^m\lambda)(\mathcal{D}_m\lambda)+(\mathcal{D}^m\omega)(\mathcal{D}_m\omega) \right]\nn\\
&&-\frac{1}{4}(\mathcal{D}^m\phi)(\mathcal{D}_m\phi), \eea
and the scalar field satisfying an equation of the form,
\be\label{laplace3}
\mathcal{D}^a\mathcal{D}_a \phi- (\mathcal{R}/8)\phi=\kappa_2^4 \phi,
\ee 
where 
\bea \kappa_2^4&=&\frac{1}{16}\lambda^{-2}\left[(\mathcal{D}^m\lambda)(\mathcal{D}_m\lambda)+(\mathcal{D}^m\omega)(\mathcal{D}_m\omega) \right]\nn\\
&&-\frac{1}{4}(\mathcal{D}^m\phi)(\mathcal{D}_m\phi). \eea

The notion of multipole moments depends on the asymptotic behavior of the potentials and in particular on whether the potentials are smooth functions at infinity. The $\kappa$ coefficients for scalar-tensor theory turn out to have the same form as in GR, with the addition of first derivatives of a harmonic function. Therefore, as in GR, the smoothness of the coefficients in the differential equations ensures the smoothness of the solutions/potentials and the definition of multipole moments is as meaningful as it is in GR. Additionally, the recursive algorithm for calculating them remains the same as in GR. 

It is worth mentioning that the asymptotic value of the scalar, $\phi_\infty$, will in principle appear in the potential for the scalar moments. In order to have the desired $1/r$ falloff at infinity, one needs to shift the scalar by $\phi_\infty$, so as to remove the constant term. This can be done without loss of generality (note that in vacuum the scalar effectively enjoys shift symmetry).

\section{Multipole moments formalism for stationary and axisymmetric spacetimes}
\label{sec:moments2}

A spacetime is stationary and axisymmetric if it admits a timelike Killing vector field $\xi^a$ and a spacelike Killing vector field $\eta^a$ that has closed integral curves. The actions of these symmetries should also commute, i.e., $\eta^a\nabla_a\xi^b-\xi^a\nabla_a\eta^b=0$. The condition for the 2-planes that are orthogonal to the two Killing vectors to be integrable is \cite{stephani,Wald:1984cw}
\be \epsilon_{abcd}\eta^a\xi^b\nabla^d\xi^c = 0 = \epsilon_{abcd}\xi^a\eta^b\nabla^d\eta^c .\ee
This condition can also be written in terms of the Ricci tensor in the form \cite{stephani}
\be
\label{condition}
\xi^d R_{d[a}\xi_b\eta_{c]} = 0 = \eta^d R_{d[a}\xi_b\eta_{c]}. \ee 
In GR this condition is satisfied in vacuum. As a consequence, the line element of stationary and axisymmetric vacuum spacetimes can take the Weyl-Papapetrou form \cite{papapetrou} 
\be ds^2=-f(dt-wd\varphi)^2+f^{-1}\left[e^{2\gamma}\left(d\rho^2+dz^2\right)+\rho^2d\varphi^2\right] \label{weyl},\ee  
without loss of generality, where the metric functions depend only on the coordinates $(\rho,z)$. 

\subsection{Axisymmetric spacetimes in scalar-tensor theory}

In the case of scalar-tensor theories, one can use the vacuum field equations in the Einstein frame, Eq.~(\ref{feq1}), in order to show that condition (\ref{condition}) is satisfied.  By virtue of the assumption that the scalar field obeys the symmetries of the metric, i.e., $\xi^a\nabla_a\phi=0$ and $\eta^a\nabla_a\phi=0$, one has that $\xi^a R_{ab}=\eta^a R_{ab}=0$. This implies that the conditions for integrability are satisfied and the line element for a stationary and axisymmetric spacetime in scalar tensor theory can be written in the Weyl-Papapetrou form, Eq.~(\ref{weyl}), without any loss of generality.

Interestingly enough, one can write the field equations for the Weyl-Papapetrou metric in the Einstein frame in the same form as they are in GR. The field equations in terms of the Ricci tensor, $R_{ab} = 2 \partial_a\phi\partial_b\phi$, can be written in terms of the Einstein tensor as 
\[G_{ab}\equiv R_{ab}-\frac{1}{2}Rg_{ab} = 2 \left(\nabla_a\phi\nabla_b\phi-\frac{1}{2}(\nabla^c\phi\nabla_c \phi) g_{ab}\right),\]
which leads to the set of equations:
\be f \bar{\nabla}^2 f = \bar{\nabla} f \cdot \bar{\nabla} f -\rho^{-2} f^4 \bar{\nabla}w \cdot \bar{\nabla}w, \ee
\be \bar{\nabla}\cdot\left(\rho^{-2}f^2\bar{\nabla}w\right)=0,\ee
where $\bar{\nabla}$ is the gradient in cylindrical flat coordinates. Using the identity $f^{-2}\bar{\nabla}w=-\rho^{-1}\hat{n}\times\bar{\nabla}w$, where $w$ is a function independent of $\varphi$ and $\hat{n}$ is a unit vector in the azimuthal direction, the above equation can be rewritten as 
\be f \bar{\nabla}^2 f = \bar{\nabla} f \cdot \bar{\nabla} f -\bar{\nabla}\omega \cdot \bar{\nabla}\omega, \ee
\be \bar{\nabla}\cdot\left(f^{-2}\bar{\nabla}\omega\right)=0.\ee
Thus there is no contribution of the scalar field and these two equations are essentially the Ernst equation \cite{Ernst1}
\be (\mathcal{R}\mathcal{E})\bar{\nabla}^2\mathcal{E}=\bar{\nabla}\mathcal{E}\cdot \bar{\nabla}\mathcal{E},\ee
where $\mathcal{E}=f+i\omega$. They are accompanied by two more equations for $\gamma$, which are 
\be \frac{\p\gamma}{\p \rho}=\left(\frac{\p\gamma}{\p \rho}\right)_{GR}+\rho \left[\left(\frac{\p\phi}{\p \rho}\right)^2-\left(\frac{\p\phi}{\p z}\right)^2\right], \label{gamma1}\ee
\be \label{gamma2} 
\frac{\p\gamma}{\p z}=\left(\frac{\p\gamma}{\p z}\right)_{GR}+2 \rho \left(\frac{\p\phi}{\p \rho}\right)\left(\frac{\p\phi}{\p z}\right),\ee
where,
\bea \left(\frac{\p\gamma}{\p \rho}\right)_{GR}&=&\frac{\rho}{4f^2}\left[\left(\frac{\p f}{\p\rho}\right)^2-\left(\frac{\p f}{\p z}\right)^2\right]\nn\\
&&-\frac{f^2}{4\rho}\left[\left(\frac{\p w}{\p\rho}\right)^2-\left(\frac{\p w}{\p z}\right)^2\right],\\
\left(\frac{\p\gamma}{\p z}\right)_{GR}&=&\frac{1}{2}\left[\frac{\rho}{f^2}\frac{\p f}{\p\rho}\frac{\p f}{\p z}-\frac{f^2}{\rho}\frac{\p w}{\p\rho}\frac{\p w}{\p z}\right],
\eea
and one more equation for the scalar field,
\be \bar{\nabla}^2\phi=0. \ee
We should note here that  $\omega$ is the scalar twist of the timelike Killing vector.
Finally, the Ernst equation can also be written in terms of a secondary Ernst potential which is defined as
\be
\label{secondary}
\zeta=\frac{1+\mathcal{E}}{1-\mathcal{E}}\,.
\ee
The field equation for $\zeta$ is,
\be 
\label{eqzeta}
(\zeta\zeta^*-1)\bar{\nabla}^2\zeta=2\zeta^*\bar{\nabla}\zeta\cdot\bar{\nabla}\zeta. \ee
These results are not entirely new, and, in a different context, one can find them in the literature (for example see \cite{Heusler} and more recently \cite{Astorino14}). 

In summary, for stationary and axisymmetric spacetimes in scalar-tensor theory, the field equations can be formulated in terms of an Ernst potential that satisfies the same equation as in GR, a set of integrability conditions for the metric function $\gamma$ that has the same form as in GR with the addition of extra terms that depend on the scalar field, and a Laplace equation for the scalar field.

\subsubsection{General spherically symmetric scalar-tensor solution by way of Ernst potential}

As an example, we can straightforwardly calculate the general spherically symmetric solution in scalar-tensor theory using the Ernst potential formalism. It is shown in Ref.~\cite{Ernst1} that,  if one makes the ansatz
\be \zeta=-e^{ia}\coth\Psi, \ee 
then eq.~(\ref{eqzeta})
becomes simply a Laplacian for the field $\Psi$, i.e., $\nabla^2\Psi=0$. Thus, one can map any ``electrostatic'' solution to a GR solution. Some of the solutions produced this way, belong to the classes of the static Weyl solutions (for $a=0$) and the Papapetrou solutions (for $a=\pi/2$) \cite{stephani}. Indeed, the Schwarzschild solution corresponds to  the potential of a rod along the $z$ axis with a uniform charge distribution \cite{Ernst1}, {\em i.e.}
\begin{equation}
\Psi=\frac{m}{2l} \log \left(\frac{\sqrt{(l+z)^2+\rho ^2}-l-z}{\sqrt{(l-z)^2+\rho
   ^2}+l-z}\right)\,, 
\end{equation}   
where $2l$ is the length of the rod and $m$ is the total ``charge.''

We can follow the same process for generating solutions in scalar-tensor theory.
Since we are interested in the solution that corresponds to the monopole of the gravitational field and the monopole of the scalar field, we can assume for the Ernst potential the same form as the one in GR. But now we need to make a choice for the scalar field as well. A reasonable choice seems to be to choose a
similar potential (which also satisfies a Laplace equation), as we would like the scalar configuration to be adapted to the matter configuration,
\begin{equation}
\label{scalaransatz}
\phi=\frac{w_A}{2l} \log \left(\frac{\sqrt{(l+z)^2+\rho ^2}-l-z}{\sqrt{(l-z)^2+\rho^2}+l-z}\right)\,,
\end{equation}   
where $w_A$ is the scalar charge and the geometry of the charge distribution is adapted to that of the gravity part.  
   
From $\Psi$ one can calculate the potential $\zeta$, and from that the Ernst potential, $\mathcal{E}=(\zeta-1)(\zeta+1)$. In this case $\mathcal{E}$ takes the form 
\be \mathcal{E}=\left(\frac{x-1}{x+1}\right)^{\frac{m}{l}}, \ee
where we have used the prolate spheroidal coordinates adapted to the rod of length $2l$, which are defined as
\be  \rho=l\sqrt{x^2 -1}\sqrt{1-y^2}, \;\;\; z=lxy, \ee  
and the fact that the solution is static having thus $a=0$. Since the Ernst potential is real, there is no rotation part in the geometry, therefore $\omega=w=0$ and $f=\left[(x-1)/(x+1)\right]^{\frac{m}{l}}$. Thus the only remaining unknown function of the metric is the function $\gamma$ which we can calculate from Eqs. (\ref{gamma1}) and (\ref{gamma2}). These equations are integrated to give the result
\be e^{2\gamma}=-C \left(\frac{x^2-1}{x^2-y^2}\right)^{\frac{m^2+w_A^2}{l^2}}, \ee
where $C$ is an integration constant. One can set $C=-1$ by a suitable coordinate rescaling and without loss of generality. The sign of $C$ is uniquely determined by the metric signature. 

So far we have generated a 3-parameter solution, as $m$, $l$, and $w_A$ are independent quantities. Since we are looking for a static, spherically symmetric solution, we expect to have just two charges, the mass and the scalar charge. Hence, we need to impose a constraint between the three parameters (recall that the formalism generically generates axisymmetric solutions; extra assumptions are needed in order to obtain spherically symmetric solutions). With a bit of foresight, we will impose
$m^2+w_A^2=l^2$.

The metric given in Weyl-Papapetrou coordinates then takes the form,
\bea 
\label{wpmetric}
ds^2&=&-\left(\frac{R_- +R_+ -2l}{R_-+R_++2l}\right)^{\frac{m}{l}}dt^2+\left(\frac{R_- +R_+ -2l}{R_-+R_++2l}\right)^{-\frac{m}{l}}\nn\\
            &&\times\left[\left(\frac{1}{2} +\frac{z^2+\rho^2-l^2}{2R_+ R_-}\right)(d\rho^2+dz^2)+\rho^2 d\varphi^2\right],\quad
\eea
where
\bea
R_+\equiv \sqrt{(l-z)^2+\rho^2}\,,\\
R_-\equiv\sqrt{(l+z)^2+\rho^2}\,.
\eea
This metric, together with the scalar profile given in Eq.~(\ref{scalaransatz}), constitutes the full solution.

 If, instead, one performs the coordinate transformation, $x=(r-l)/l$, and $y=\cos\theta$, one obtains the metric 
\bea ds^2&=&-\left(1-\frac{2l}{r}\right)^{\frac{m}{l}} dt^2+\left(1-\frac{2l}{r}\right)^{-\frac{m}{l}} dr^2\nn\\
&&+\left(1-\frac{2l}{r}\right)^{1-\frac{m}{l}} r^2(d\theta^2+\sin^2\theta d\varphi^2). \eea
In these coordinates, the scalar field potential becomes 
\be
\phi=\frac{w_A}{2l}\log\left(1-\frac{2l}{r}\right)\,.
\ee 
This is indeed the known static, spherically symmetric solution in scalar-tensor theory \cite{Just1959ZNat, Damour}.\footnote{In the notation of Ref.~\cite{Damour}, the length of the rod, $2l$, corresponds to the parameter $a$ of the metric, twice the charge, $2m$, corresponds to the parameter $b$ of the metric, and finally the charge $w_A$ corresponds to the parameter $d$ for the scalar field. This explains why we have chosen to set $m^2+w_A^2=l^2$ which corresponds to the identity $a^2-b^2=4d^2$ in Ref.~\cite{Damour}. The comparison also reveals that $m$ is the Einstein frame mass.} For $w_A=0$ the scalar configuration becomes trivial and the metric reduces to the Schwarzschild solution. Note that, even though it might not be evident in these coordinates, when $w_A\neq 0$ the metric is regular everywhere apart from $r=2l$. Moreover, $r=2l$ is actually a curvature singularity and the scalar diverges there as well ($r=2l$ corresponds to the vanishing areal radius). The requirement that there be a horizon covering the singularity imposes $w_A=0$, in agreement with the no-hair theorem for scalar-tensor theory \cite{Hawking:1972qk,Sotiriou:2011dz}.

\subsection{Axisymmetric multipole moments}

Following Ref.~\cite{fodor:2252}, one can calculate the moments of stationary and axisymmetric spacetimes using the Ernst potential. The Ernst potential is given as $\mathcal{E}=f+i\omega$, where $f$ is the norm of the timelike Killing vector and $\omega$ is the twist potential of the timelike Killing vector. Thus one can use the Ernst potential to define the potentials that will give the moments. It is actually convenient to define
\be \xi=\frac{1-\mathcal{E}}{1+\mathcal{E}}\,. \ee
instead and use it for the definition of the moment. In terms of the moment potentials 
\bea \Phi_M &=&\frac{(1-\lambda^2-\omega^2)}{(1+2\lambda+\lambda^2+\omega^2)}\,,\\
         \Phi_J &=& \frac{-2\omega}{(1+2\lambda+\lambda^2+\omega^2)}\,,
 \eea         
 one has $\xi= \Phi_M+ i \Phi_J$. 
Notice that $\xi$ is actually the inverse of $\zeta$, that was defined in Eq.~(\ref{secondary}) \cite{fodor:2252}. However, $\xi$ and $\zeta$ actually satisfy the same equation, {\em i.e.}
\be(\xi\xi^*-1)\bar{\nabla}^2\xi=2\xi^*\bar{\nabla}\xi\cdot\bar{\nabla}\xi. \ee

The moments will be given by the recursive algorithm,
\bea  P &=& \tilde{\xi},\nn\\
         P_a &=& \bar{D}_a P,\\
         P_{a_1 \ldots a_{s+1}} &=& \mathbf{\mathcal{C}}\left[\bar{D}_{a_1}P_{a_2 \ldots a_{s+1}}-\frac{s}{2}(2s-1)\bar{R}_{a_1a_2}P_{a_3 \ldots a_{s+1}}\right].\nn
  \eea    
The difference from the purely stationary case is that, due to extra rotational symmetry, the moments will now be some multiples of the symmetric trace free outer product of the axis vector and correspond to only one component of that tensor. Hence, they will be scalar quantities,
\be P_n=\left.\frac{1}{n!}P^{(n)}_{i_1\ldots i_n} n^{i_1}\ldots n^{i_n}\right |_{\Lambda} = \left. \frac{1}{n!}P^{(n)}_{2\ldots 2}\right |_{\Lambda}. \ee

From the metric in the Weyl-Papapetrou coordinates (\ref{weyl}) one gets to the line element of the three-manifold, 
\be   ds_{(3)}^2= e^{2\gamma(\rho,z)}(d\rho^2+dz^2)+\rho^2 d\varphi^2\,. \ee
After the coordinate transformation $\bar{\rho}=\rho/(\rho^2+z^2)$,   $\bar{z}=z/(\rho^2+z^2)$ that maps infinity to a point $\Lambda$ (the center of the coordinate system), this line element takes the form
 \be   ds_{(3)}^2=\frac{1}{\bar{r}^4}\left[ e^{2\gamma(\bar{\rho},\bar{z})}(d\bar{\rho}^2+d\bar{z}^2)+\bar{\rho}^2 d\varphi^2\right], \ee
where $\bar{r}^2=\bar{\rho}^2+\bar{z}^2$. Thus the conformal metric of the three-manifold with the center of coordinates at infinity is given as $\tilde{h}_{ij}=\Omega^2 h_{ij}=\bar{r}^4 h_{ij}$, where $\Omega=\bar{r}^2$ is the conformal factor. 

In these coordinates, the field equations for $\xi$ and $\phi$ become
\bea (\bar{r}^2\tilde{\xi}\tilde{\xi}^*-1)\nabla^2\tilde{\xi}&=&2\tilde{\xi}^*[\bar{r}^2(\nabla\tilde{\xi})^2 + 2\bar{r}\tilde{\xi}\nabla\tilde{\xi} \cdot \nabla\bar{r} + \tilde{\xi}^2]\,,\quad\\
 \nabla^2\tilde{\phi}&=&0,\eea
where $\tilde{\xi}=(1/\bar{r})\xi$, $\tilde{\phi}=(1/\bar{r})\phi$, and the derivatives are the flat derivative operators at infinity in cylindrical coordinates. The Ricci tensor of the three-manifold at conformal infinity can be expressed with respect to the fields $\tilde{\xi}$, and $\tilde{\phi}$ as,
\be \tilde{R}_{ij}=\frac{1}{D^2}(G_iG_j^*+G_i^*G_j)+\frac{2}{\bar{r}^2}G_i^{\phi}G_j^{\phi}, \ee
where, as in GR, $D=\bar{r}^2\tilde{\xi}\tilde{\xi}^*-1$, $G_1=\bar{z}\tilde{\xi}_{,1}-\bar{\rho}\tilde{\xi}_{,2}$, $G_2=\bar{\rho}\tilde{\xi}_{,1}+\bar{z}\tilde{\xi}_{,2}+\tilde{\xi}$, $G_3=0$, and additionally for scalar-tensor theory, $G_1^{\phi}=\bar{z}(\bar{r}\tilde{\phi})_{,1}-\bar{\rho}(\bar{r}\tilde{\phi})_{,2}$, $G_2^{\phi}=\bar{\rho}(\bar{r}\tilde{\phi})_{,1}+\bar{z}(\bar{r}\tilde{\phi})_{,2}$, and $G_3^{\phi}=0$. 
Finally the derivatives of the $\gamma$ function are given as
\be \gamma_{,1}=\frac{\bar{\rho}}{2}(\tilde{R}_{11}-\tilde{R}_{22}),\; \gamma_{,2}=\bar{\rho}\tilde{R}_{12}. \ee

As in the case of GR, the potential $\tilde{\xi}$ can be expressed as a series expansion around infinity of the form
\be \tilde{\xi}=\sum_{i,j=0}^{\infty} a_{ij}\bar{\rho}^i\bar{z}^j, \ee
where the coefficients $a_{ij}$ can be expressed with respect to the coefficients $a_{0j}=m_j$ of the expansion of $\tilde{\xi}$ along the axis of symmetry, $\tilde{\xi}(\bar{\rho}=0)=\sum_{j=0}^{\infty} m_j\bar{z}^j$. The recursive relation that connects the various $a_{ij}$ as it can be evaluated from the field equation for $\tilde{\xi}$ is 
\bw
 \bea (r+2)^2 a_{r+2,s}&=&-(s+2)(s+1)a_{r,s+2}+\!\!\!\!\!\!\!\!\!\!\!\!  \sum_{
\begin{array}{c}
k+m+p=r\\
l+n+q=s   
\end{array}
}   \!\!\!\!\!\! \!\!\!\!\!\! a_{kl}a^*_{mn}\times\left[\begin{array}{l} a_{pq}(p^2+q^2-4p-5q-2pk-2ql-2) \\+ a_{p+2,q-2}(p+2)(p+2-2k)\\ +a_{p-2,q+2}(q+2)(q+1-2l)\end{array}\right].  \eea
\ew
If $i$ is an odd number, then $a_{ij}=0$. In a similar way, one can express $\tilde{\phi}$ as a series expansion at infinity of the form,
\be \tilde{\phi}=\sum_{i,j=0}^{\infty} b_{ij}\bar{\rho}^i\bar{z}^j.\ee
One can see from the field equation for $\tilde{\phi}$ that the coefficients $b_{ij}$ satisfy the recursive relation,
\be b_{i+2,j}=-\frac{(j+2)(j+1)}{(i+2)^2} b_{i,j+2}, \ee
and that $b_{1,j}=0$. If one further assumes reflection symmetry about the equatorial plane, which is a reasonable assumption for quiescent astrophysical objects, then the  coefficients of odd powers of $\bar{z}$ should be $b_{i,2j+1}=0$. Again all the $b_{ij}$ coefficients can be calculated from the expansion of $\tilde{\phi}$ along the symmetry axis, $\tilde{\phi}(\bar{\rho}=0)=\sum_{j=0}^{\infty} w_j\bar{z}^j$.

Following Ref.~\cite{fodor:2252}, the moments for both the gravitational field and the scalar field can be calculated by following the algorithm for calculating the polynomials $S$: 
\begin{itemize}
\item Set $S^{1(0)}_0 =\tilde{\xi}$ and $S^{2(0)}_0 =\tilde{\phi}$,
\item Set $S^{i(1)}_0=\frac{\p}{\p\bar{z}}S^{i(0)}_0$, and $S^{i(1)}_1=\frac{\p}{\p\bar{\rho}}S^{i(0)}_0$,
\item Let 
\bw
\bea S^{i(n)}_a &=&\frac{1}{n}\left\{ a\frac{\p}{\p\bar{\rho}}S^{i(n-1)}_{a-1}+(n-1)\frac{\p}{\p\bar{z}}S^{i(n-1)}_{a}\right.+a\left[(a+1-2n)\gamma_{,1}-\frac{a-1}{\bar{\rho}}\right]S^{i(n-1)}_{a-1}+(a-n)(a+n-1)\gamma_{,2}S^{i(n-1)}_{a}\nn\\
&&+a(a-1)\gamma_{,2}S^{i(n-1)}_{a-2}+(n-a)(n-a-1)\left(\gamma_{,1}-\frac{1}{\bar{\rho}}\right)S^{i(n-1)}_{a+1}-\left(n-\frac{3}{2}\right)[a(a-1)\tilde{R}_{11}S^{i(n-2)}_{a-2}\nn\\
&&+2a(n-a)\tilde{R}_{12}S^{i(n-2)}_{a-1} \left.+(n-a)(n-a-1)\tilde{R}_{22}S^{i(n-2)}_{a}]\right\}
\eea
\ew
where $0\leq a\leq n$, and the superscript $i$ refers to either the gravity potential for $i=1$ or the scalar potential for $i=2$. 

\item The moments are then given by 
\be P^i_n=\left.\frac{1}{(2n-1)!!}S^{i(n)}_{0}\right|_{\Lambda} \ee

\end{itemize}

Following this algorithm, the first few moments are
\bw
\bea  P^g_0&=&m_0,\;   P^g_1=m_1,\; P^g_2=m_2-\frac{1}{3}m_0w_0^2,\;  P^g_3=m_3-\frac{3}{5}m_1w_0^2,\nn\\
         P^g_4&=&m_4-\frac{1}{7}(m_0m_2-m_1^2)m_0^*-\frac{30}{105}(3m_2w_0+m_0w_2)w_0+\frac{1}{105}(8m_0m_0^*+19w_0^2)m_0w_0^2,\nn\\
         P^g_5&=&m_5-\frac{1}{3}(m_0m_3-m_1m_2)m_0^*-\frac{1}{21}(m_2m_0-m_1^2)m_1^*+\frac{2}{63}(m_0^2m_1^*-35m_3)w_0^2+\frac{5}{63}(2m_0m_0^*w_0+5w_0^3-6w_2)m_1w_0,\nn\\
         P^g_6&=&  m_6+ \frac{1}{33} (m_2m_0 -m_1^2) m_0 \left(m_0^*\right)^2 -\frac{5}{231} m_2^* (m_0 m_2-m_1^2) -\frac{4}{33} (m_0 m_3-m_1 m_2) m_1^*-\frac{8}{33} (m_1 m_3-m_2^2 ) m_0^*\nn\\
         &&-\frac{6}{11} (m_0 m_4-m_1 m_3) m_0^*+\frac{1}{3465}(-389 m_0 w_0^6-392 m_0^2 m_0^* w_0^4+1470 m_0 w_2 w_0^3-72 m_0^3 \left(m_0^*\right)^2 w_0^2\nn\\
          &&-495 m_1^2 m_0^* w_0^2+252 m_0 m_1 m_1^* w_0^2-630 m_0 w_4 w_0+720 m_0^2 m_0^* w_2 w_0-450 m_0 w_2^2\nn\\
          &&-4725 m_4 w_0^2+60 m_2^* m_0^2 w_0^2+15 m_2 \left(89 m_0m_0^* w_0^2+155 w_0^4-150 w_2 w_0\right)),\nn\eea
 and,
 \bea
         P^{\phi}_0&=&w_0,\;  P^{\phi}_1=0, \;  P^{\phi}_2=w_2-\frac{1}{3}w_0(m_0m_0^*+w_0^2),\;  P^{\phi}_3=-\frac{1}{5}(m_1m_0^*+m_0m_1^*)w_0=0,\nn\\
         P^{\phi}_4&=&w_4+ \frac{1}{105} \left(38 m_0 m_0^* w_0^3-3 \left(-3 m_0^2 \left(m_0^*\right)^2+5 m_2 m_0^*+6 m_1 m_1^*+5 m_0 m_2^*\right) w_0\right.\nn\\
         &&\left.-90 m_0m_0^* w_2+19 w_0^5-120 w_2 w_0^2\right)  ,\nn\\
         P^{\phi}_5&=&  \frac{1}{63} \left(-7 m_3 m_0^* w_0+m_1^* \left(2 m_0 \left(3 m_0 m_0^* w_0+8 w_0^3-15 w_2\right)-9 m_2 w_0\right)\right.\nn\\
         &&\left.+m_1 \left(2 m_0^* \left(3 m_0m_0^* w_0+8 w_0^3-15 w_2\right)-9 m_2^* w_0\right)-7 m_0 m_3^* w_0\right) =0,\nn\\
         P^{\phi}_6&=&  w_6  +  \frac{1}{3465}(-1167 m_0 m_0^* w_0^5+\left(-677 m_0^2 \left(m_0^*\right)^2+735 m_2 m_0^*+882 m_1 m_1^*+735 m_0 m_2^*\right) w_0^3\nn\\
         &&+765 w_0^2 \left(8 m_0 m_0^* w_2-7 w_4\right)-15 w_0 \left(5 m_0^3 \left(m_0^*\right)^3-3 m_1^2 \left(m_0^*\right)^2-18 m_0 m_2 \left(m_0^*\right)^2+21 m_4 m_0^*\right.-18 m_0^2 m_2^* m_0^*\nn\\
         &&\left.-3 m_0^2 \left(m_1^*\right)^2+28 m_3 m_1^*+30 m_2 m_2^*+m_1 \left(28 m_3^*-30
   m_0 m_0^* m_1^*\right)+21 m_0 m_4^*+180 w_2^2\right)\nn\\
   &&+45 \left(-5 \left(-7 m_0^2 \left(m_0^*\right)^2 w_2+5 m_2 m_0* w_2+6 m_1m_1^* w_2+m_0 \left(21 m_0^* w_4+5 m_2^* w_2\right)\right)\right)-389 w_0^7+3795 w_2 w_0^4)\,.\nn  
\eea
\ew
Our assumption that the configuration exhibits reflection symmetry about the equatorial plane implies that the odd multipole coefficients $m_{2n+1}$ are imaginary, while the even coefficients $m_{2n}$ are real.  This means that all the combinations that appear as $m+m^*$ are equal to zero.  Hence, all odd scalar moments vanish. It is straightforward, though computationally more challenging, to calculate the moments without assuming reflection symmetry. We give the result in the Appendix. 

As an example and a sanity check, we can calculate the moments of the spherically symmetric solution in scalar-tensor theory that we presented in the previous section. By definition, the  gravitational monopole is equal to $m$, the scalar monopole is equal to $w_A$, all higher moments are equal to zero, and this is what we expect to find. The Ernst potential is  

\be \mathcal{E}=\left(\frac{-2l+\sqrt{\rho^2+(z-l)^2}+\sqrt{\rho^2+(z+l)^2}}{2l+\sqrt{\rho^2+(z-l)^2}+\sqrt{\rho^2+(z+l)^2}}\right)^{\frac{m}{l}}, \ee
from which one can calculate 
\be
\tilde{\xi}=(1/\bar{r})\xi|_{\!\!\!\tiny{\begin{array}{l}\rho\rightarrow\bar{\rho}\\ z\rightarrow\bar{z} \end{array}}}
\ee
 and then expand it in powers of $\bar{z}$ along the axis of symmetry $\bar{\rho}=0$. Thus the first few coefficients $m_i$ are,
\bea m_0&=&m,\nn\\
        m_1&=&0,\nn\\
        m_2&=&\frac{1}{3}(l^2-m^2)m,\nn\\
        m_3&=&0,\nn\\
        m_4&=&\frac{1}{15}(3l^4m-5l^2m^3+2m^5),\\
        m_5&=&0,\nn\\
        m_6&=&\frac{1}{315}(l^2-m^2)(45l^4-53l^2m^2+17m^4)m,\nn\\
        &\vdots&\nn
\eea
Similarly, for the scalar field $\tilde{\phi} =\left.(1/\bar{r})\phi\right|_{\!\!\!\tiny{\begin{array}{l}\rho\rightarrow\bar{\rho}\\ z\rightarrow\bar{z} \end{array}}}$, the expansion gives the coefficients, 
\bea w_0&=&-w_A,\nn\\
        w_1&=&0,\nn\\
        w_2&=&-\frac{1}{3}l^2w_A,\nn\\
        w_3&=&0,\nn\\
        w_4&=&-\frac{1}{5}l^4w_A,\\
        w_5&=&0,\nn\\
        w_6&=&-\frac{1}{7}l^6w_A,\nn\\
        &\vdots&\nn
\eea
By substituting these quantities in the expressions for the moments and taking into account the fact that $l^2=w_A^2+m^2$, the above mentioned result is recovered, i.e., all moments vanish apart from the gravitational monopole and the scalar monopole.

\section{Conclusions}

We have developed the formalism for defining the multipole moments for stationary, asymptotically flat spacetimes in scalar-tensor theories with one massless scalar field. We have also demonstrated how
 one can extend and apply the Ernst formulation of GR to scalar-tensor theories in order to produce stationary and axisymmetric solutions. Taking advantage of the Ernst formulation of the field equations we have outlined the process and obtained an algorithm for calculating the multipole moments of stationary and axisymmetric solutions in scalar-tensor theory from the Ernst potential.
 
 It has proved much more convenient to define the moments in the Einstein frame. This is because the Einstein frame scalar field satisfies a standard Laplace equation for a suitable coordinate choice. However, it is worth stressing that one could have defined the moments by working directly in the Jordan frame and attempting to reformulate the equations in order to find appropriate potentials that can be used for the definition. Since the Jordan frame metric and scalar configuration are uniquely related with their Einstein frame counterparts, in practice there is no need to define the moments directly in the Jordan frame in order to characterize a solution. Any observable can be expressed in terms of the Einstein frame moments defined above.
 
 We expect our results to be an important stepping stone toward using compact objects as probes for deviations from GR. For example, our results can  be used to calculate  the multipole moments of fluid configurations, employing a numerical scheme as the one developed in Ref.~\cite{Doneva2013PhRvD}. As already mentioned, it has been shown in Refs.~\cite{PaniBerti2014PhRvD,Doneva2014PhRvD} that there is  degeneracy between GR and scalar-tensor theory with respect to the {\it I-Love-Q} relations of neutron stars.\footnote{We should also note  that in Ref.~\cite{Doneva2014PhRvD} a calculation was presented for the quadrupole of the spacetime. This result is in full agreement with our result for the mass quadrupole.} One could hope to break this degeneracy by exploring the relations between the higher order moments and the lower order moments in the two theories. If such a distinction is possible, it will open a possibility for testing deviations from GR.

Clearly, our main motivation for developing the formalism for defining and calculating the multipole moments has been to apply it to fluid configurations describing neutron or quark stars in scalar-tensor gravity. However, other application could include scalarized black holes as well. Isolated, stationary, asymptotically flat black holes in scalar-tensor gravity are expected to be identical to those of GR \cite{Hawking:1972qk,Sotiriou:2011dz}. On the other hand, it has been recently pointed out that, in the presence of some matter configuration around them, black holes in scalar-tensor gravity could develop a scalar charge \cite{Sotiriou2013PRL,Cardoso:2013opa}. 
As long as the matter component is restricted to a finite region around the black hole, then a vacuum solution and a massless scalar field configuration can be defined for the exterior region and the multipole moments can be evaluated in the manner that is presented here.\footnote{It is worth mentioning that black hole solutions with non-trivial scalar field configurations have been also discussed in \cite{Yazadjiev1,Yazadjiev2}, but these solutions possess electromagnetic charges and therefore the methods presented here would not apply to them.}

We would like to close with a comment regarding the case where the scalar field has a nonvanishing potential $V(\phi)$. The presence of a potential would result in an extra term on the right hand side of Eq.~(\ref{feq1}) that would be proportional to the metric times the potential $V(\phi)$ and a source term for the scalar field in Eq.~(\ref{feq2}) that would be proportional to the derivative of the potential, $dV/d\phi$. After adding these terms, the right hand side of Eq.~(\ref{twistcond}) is still vanishing. So, the presence of a potential  does not prevent the twist from being curl-free, which would have been a major technical obstruction in defining the moments. Additionally we should point out that in the case of axisymmetric spacetimes, the conditions (\ref{condition}) will still be satisfied and therefore the line element can be written in the Weyl-Papapetrou (\ref{weyl}) form. However, one does need to modify the definition of the potentials that are used to calculate the moments, as the field equations have clearly changed. In conclusion, the formalism presented here is not directly applicable to theories with a potential, but it seems likely that it can be extended to include this case.

\acknowledgments

We are grateful to E. Berti, H.~O. da Silva, and S. S. Yazadjiev for a critical reading of the manuscript and helpful comments.
G.P. would like to thank S. S. Yazadjiev for some very useful discussions on the Ernst formulation of scalar-tensor theories, and K. D. Kokkotas for his hospitality at the University of T\"ubingen.
The research leading to these results has received funding  from the European Research Council under the European Union's Seventh Framework Programme (FP7/2007-2013) / ERC grant agreement n. 306425 ``Challenging General Relativity''.  

\appendix

\section{Axisymmetric moments without reflection symmetry}

If we relax the requirement that the scalar field exhibit reflection symmetry about the  equatorial plane, then the odd terms in the expansion of the scalar field along the $z$ axis, $w_{2j+1}$, will not be equal to zero. This will lead to the following extra contributions to the axisymmetric multipole moments,
\bw

\bea 
\Delta P^g_4&=&-\frac{6}{35} w_1 \left(4 m_1 w_0+m_0 w_1\right)\nn\\
\Delta P^g_5&=&\frac{1}{63} \left(-14 m_0 w_0 w_3-2 w_1 \left(30 m_2 w_0+9 m_1 w_1+m_0 \left(-6 m_0 m_0^* w_0-16 w_0^3+9 w_2\right)\right)\right)\nn\\
\Delta P^g_6&=&\frac{1}{1155}\left(2 w_1 \left(24 m_0^2 \left(3 m_0^* w_1+2 m_1^* w_0\right)+m_0 w_0 \left(228 m_1 m_0^*+335 w_0 w_1\right)-700 m_3 w_0-225 m_2 w_1\right.\right.\nn\\
&&\left.\left.+30 m_1 \left(20
   w_0^3-9 w_2\right)\right)-140 w_3 \left(3 m_1 w_0+2 m_0 w_1\right)\right)
   \eea
   \bea
\Delta P^{\phi}_1&=& w_1\nn\\
 \Delta P^{\phi}_3&=&  w_3-\frac{1}{5} w_1 \left(3 m_0 m_0^*+5 w_0^2\right)\nn\\
 \Delta P^{\phi}_4&=&-\frac{6}{35} w_1 \left(2 m_1 m_0^*+2 m_0 m_1^*+5 w_0 w_1\right)\nn\\
  \Delta P^{\phi}_5&=&\frac{1}{63} \left(-14 w_3 \left(5 m_0 m_0^*+6 w_0^2\right)-w_1 \left(18 \left(m_1 m_1^*+w_1^2\right)+m_0 \left(15 m_2^*-82 m_0^* w_0^2\right)-15 m_0^2 \left(m_0^*\right)^2\nn\right.\right.\\
   &&\left.\left.+15 m_2 m_0^*-57 w_0^4+108 w_2 w_0\right)+63 w_5\right)\nn\\
   \Delta P^{\phi}_6&=&\frac{1}{1155}\left(2 w_1 \left(488 m_1 m_0^* w_0^2+m_0 \left(488 m_1^* w_0^2+747 m_0^* w_1 w_0+150 m_1 \left(m_0^*\right)^2-105 m_3^*\right)\right.\right.\nn\\
   &&\left.\left.-105 m_3 m_0^*-135 m_2 m_1^*+150 m_0^2 m_0^* m_1^*-135 m_1 m_2^*+935 w_1 w_0^3-495 w_1 w_2\right)\right.\nn\\
   &&\left.-700 w_3 \left(m_1 m_0^*+m_0 m_1^*+3 w_0 w_1\right)\right)
\eea

\ew

Additionally, the mass moments coefficients $m_j$ will not split into pure real for odd $j$ and pure imaginary for even $j$ (all $m$ will be complex). As a result, the  terms that appear in the scalar field odd moments in the expressions presented in the main text will no longer vanish. The full expressions for the moments will be given by the expressions shown in the main text supplemented by the extra terms that appear in this appendix.

\bibliography{mybibliography}

\end{document}